\documentclass[11pt]{article}

\usepackage{fullpage}
\usepackage{authblk}  
\usepackage{wrapfig}

\usepackage{enumitem}
\usepackage{amsmath}
\usepackage{graphicx}
\usepackage{placeins}

\usepackage{hyperref}
\usepackage{cleveref}
\hypersetup{
    colorlinks = false,
    pdfborder = {0 0 0}
}

\usepackage{subcaption}

\usepackage{nomencl} 
\makenomenclature

\usepackage{sidecap}
\usepackage{svg}
\usepackage[version=4]{mhchem}
\usepackage{siunitx}
\usepackage{booktabs} 
\usepackage[left=0.8in, right=0.8in, top=0.8in, bottom=0.8in]{geometry}

\title{VNS Tokamak OpenMC-Serpent Validation for Medical Isotope Studies\thanks{This work has been submitted to PHYSOR 2026 - The International Conference on Physics of Reactors.}}

\author[1,2]{Christopher Ehrich}
\author[3]{Christian Bachmann}
\author[3]{Pavel Pereslavtsev}
\author[1,2]{Christian Reiter}

\affil[1]{Forschungs-Neutronenquelle Heinz Maier-Leibnitz (FRM II), Lichtenbergstraße 1, Garching bei München, Germany}
\affil[2]{Chair of Applied Nuclear Technologies, School of Engineering and Design, Technical University of Munich, Boltzmannstraße 15, Garching bei München, Germany}
\affil[3]{EUROfusion Consortium, FTD Department, Boltzmannstraße 2, Garching bei München, Germany}

\date{}

\begin{document}
\maketitle
\vspace{-4mm}
\begin{abstract}
The Volumetric Neutron Source (VNS) tokamak is a proposed fusion reactor for testing and qualification of reactor components for future use in a fusion power facility, and has potential use for radioisotope production. The VNS geometry is modeled in the Serpent and OpenMC neutronics codes. Analog neutron-photon coupled simulations are carried out to compare the model's vacuum vessel and blanket components across codes. In the vacuum vessel, neutron and photon flux maps are calculated, while in the blanket region, neutron and photon spectra, $(n,T)$, and $(n,2n)$ reaction rates are calculated and compared between models. The detector response comparisons found the following: neutron flux and $(n,T)$ reactions achieved excellent agreement, the $(n,2n)$ detector response had good agreement, and photon flux had regional discrepancies depending on Serpent tracking used. Hybrid tracking lead to a relative difference of about $20\%$ in the outboard side blanket, where as employment of delta tracking resulted in less than $1\%$ relative difference. On an HPC cluster, Serpent was found to have shorter computation time than OpenMC in neutron photon coupled simulations using both hybrid tracking and delta tracking, but longer in neutron only simulations. An exemplary radioisotope production case is presented for the demonstration of additional VNS capabilities. 
\end{abstract}

\section{Introduction}\label{sec:1}
Nuclear fusion has recently received significant political support in Germany, as it presents an attractive technology for large scale clean energy production with an abundant fuel supply. The``Aktionsplan Fusion''\cite{BMFTR2025} a measure of the Hightech Agenda Deutschland, has set a goal to build the world's first fusion power plant in Germany. The most achievable fusion fuel cycle is the \ce{D}-\ce{T} reaction, because it has largest cross section at the lowest temperature for fusion standard conditions. This reaction produces an additional neutron, \SI{17.6}{MeV}, and a \ce{He} nucleus. \SI{14.1}{MeV} of the energy is carried away by the neutron for energy production, Tritium breeding, and other applications. For future fusion energy applications, a sustainable Tritium supply is crucial. For regular operation, e.g. in large commercial fusion power plants, Tritium breeding blankets are planned. However, especially for startup, there might not be enough Tritium \cite{PEARSON20181140}. Fission reactors might help to overcome this issue. The primary \ce{T} production reactions in fusion are:
\begin{align}
    ^6\ce{Li}+n &\rightarrow ^3\ce{T}+^4\ce{He}   +\text{\SI{4.8}{MeV}} \\
    ^7\ce{Li}+n &\rightarrow ^3\ce{T} +^4\ce{He}  +n -\text{\SI{2.5}{MeV}}
\end{align}
The $^7$\ce{Li} \ce{T}-production reaction can only take place above the \SI{2.5}{MeV} threshold energy, while $^6$\ce{Li} has a much larger thermal cross section, but negatively affects neutron balance. Thus the nature of the neutron economy of a fusion reactor requires a highly optimized \ce{T} breeding blanket.

\section{Volumetric Neutron Source (VNS)}
The VNS is a Q=1 plasma tokamak, requiring four tangential \SI{120}{keV} deuteron beamlines and electron cyclotron heating to maintain a quasi steady-state fusion reaction, generating $\approx$\SI{30}{MW} of fusion power. The primary objective of VNS is to qualify the effects of high \SI{14}{MeV}-neutron flux and fluence over a large volume on reactor components, such as  Tritium breeding and shielding blankets, superconducting coils, and the vacuum vessel. A detailed description of the design concept can be found in \cite{BACHMANN2025114796, Leichtle20062025}. An important aspect that differs VNS from fusion reactors targeting power generation is the far lower Tritium consumption ($<$\SI{1}{kg / a} \cite{BACHMANN2025114796}) due to the much lower fusion power and small major radius. This could be provided for by current availability from Canadian and Korean CANDU reactors, with their combined fleets producing $\approx$\SI{2.7}{kg / a} \cite{PEARSON20181140,NI20132422}. Additionally, high-flux research reactors like FRM II, provide potential for \ce{T} production, thus fostering the \ce{T} supply chain in Europe. The VNS will test various blanket modules designed to serve three primary roles:
\begin{enumerate}[itemsep=0pt, topsep=2pt]
    \item Tritium Breeding Blanket Testing
    \item Radiation Protection Studies
    \item Medical Isotope Production 
\end{enumerate}

As Tritium breeding is foreseen in the VNS for testing purposes only, a large neutron surplus is available for additional nuclear reactions. One promising application of the VNS that should be investigated, is medical isotope production. Thanks to the large volume, high flux, and absence of a direct feedback to the fusion reaction by neutron flux, magnetic confinement fusion reactors are excellent candidates for large volume medical isotope production \cite{pereslavtsev2024potential}.

Because of the uniquely efficient $\ce{W}+\ce{TiH_2}$ shielding in the VNS geometry and the high leakage rate from additional external heating ports, there could be discrepancies in detector response estimates across codes, and it is crucial to have reliable simulation tools to assess the feasibility of the aforementioned applications in the VNS. If the reaction rates differ between codes, it follows that predictions for medical isotope production also can, and it is important to qualify the results. In this context, neutron flux, photon flux, (\textit{n,T}), (\textit{n,2n}) reaction rates, and the efficiency of high performance computer (HPC) simulation are compared in this work for a model of the VNS\cite{BACHMANN2025114796} using both Serpent 2.2 \cite{LEPPANEN2015142} and OpenMC version 0.15.2 \cite{ROMANO201590}. As a representative example, the generation of widely used $^{99}$\ce{Mo} \cite{NEA2016} is simulated in Serpent.

\section{Codes}
Two continuous-energy Monte Carlo transport codes were chosen for comparison in this study. In general, similar results should be expected since most of the underlying principles of both codes are very similar, as has been shown in recent validation work such as \cite{10682518,VALENTINE2022113197, valentine2021benchmarking}. However, the codes differ in the estimators used for detector responses, depending on the thickness and macroscopic cross section of regions \cite{LEPPANEN2017161}.
\subsection{Serpent}
Serpent is a continuous energy Monte Carlo neutron and photon transport code. Models are implemented using universe based Constructive Solid Geometry (CSG). Serpent is developed and maintained by VTT in Finland \cite{LEPPANEN2015142}, is well established in the reactor physics community, and distributed via license acquisition. Serpent is OpenMP and MPI parallelizable.
\subsection{OpenMC}
OpenMC, similar to Serpent is a continuous energy Monte Carlo neutron and photon transport code, with models implemented via CSG \cite{ROMANO201590}. OpenMC has the advantage of being a full open source and community-driven code, meaning that the program's capabilities are expanded directly by the community in addition to the developers, whereas Serpent is not open source. OpenMC is MPI and OpenMP parallelizable and was designed for HPC scalability since its inception.
\section{Model}
\begin{figure}[htbp]
\vspace{-6pt}
    \centering
    \begin{subfigure}[t]{0.45\textwidth}
        \centering
        \includegraphics[width=\textwidth]{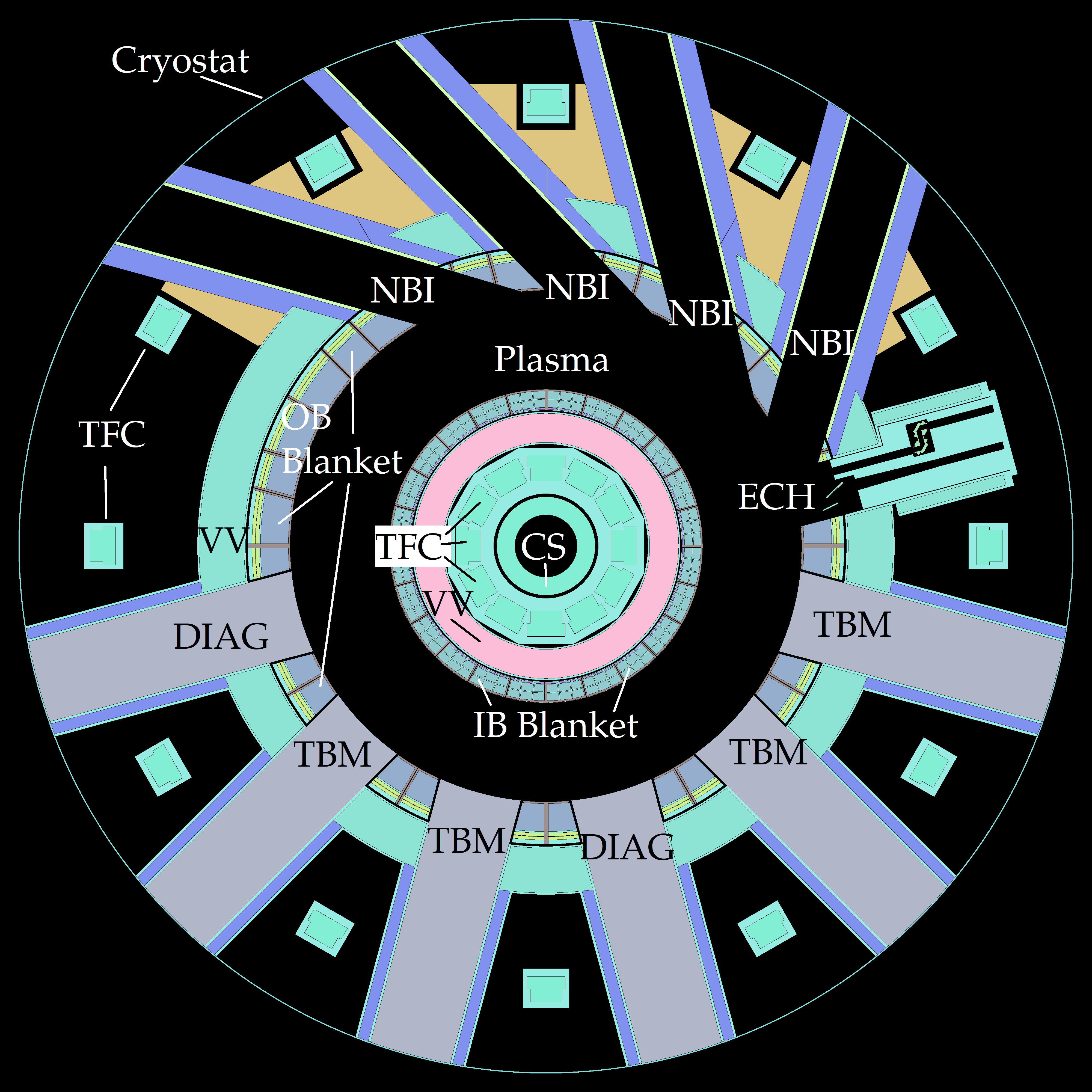}
        \caption{VNS Equatorial Geometry Cross Section}
        \label{fig:xy_geometry}
    \end{subfigure}
    \hfill
    \begin{subfigure}[t]{0.45\textwidth}
        \centering
        \includegraphics[width=\linewidth]{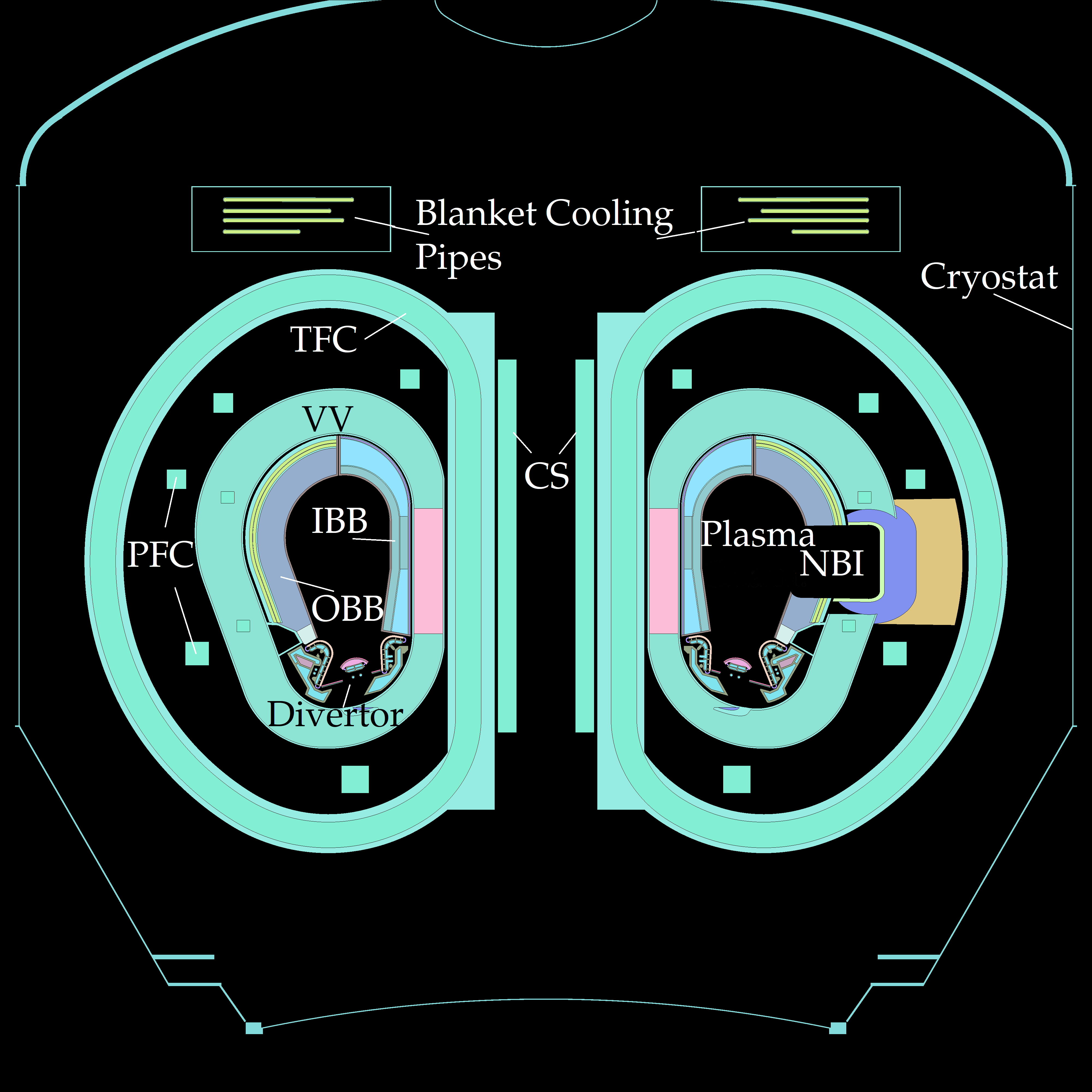}
        \caption{VNS Side Profile Geometry Cross Section}        \label{fig:xz_geometry}
    \end{subfigure}
    \label{fig: geometry}
    \caption{VNS Overview: Neutral Beam Injector (NBI), Inboard Blanket (IBB), Outboard Blanket (OBB), Test Blanket Module (TBM), Diagnostic Modules (DIAG) Plasma,, Vacuum Vessel (VV), Central Solenoid (CS), Toroidal and Poloidal Field Coils (TFC's and PFC's), Blanket Cooling Pipes, Divertor, and Cryostat}
\vspace{-6pt}
\end{figure}
The models were translated independently from an MCNP6 CSG model produced by EUROfusion \cite{Leichtle20062025}. A python script was developed, translating the MCNP model geometry, materials, and the plasma neutron source of the VNS from MCNP to Serpent. The resultant Serpent model was thoroughly checked for geometry errors such as cells and cell overlap using the built-in geometry plotter. An annotated reactor geometry details is to be found in Figure 1. The only discrepancy found in the model from the MCNP case, was the source energy distribution. In MCNP the the source neutron energy follows a Gaussian distribution, which would require modifying the source code of Serpent to implement. Instead a 20-bin weighted energy mesh following the Gaussian was used to accurately sample the energy distribution of source neutrons. The OpenMC model was also scripted from the MCNP model, but using the ``openmc\textunderscore mcnp\textunderscore adapter'' tool developed by \cite{openmc_mcnp_adapter}, for the geometry and materials. A custom script was made to translate the source mesh from MCNP to OpenMC as source translation has not been implemented in the adapter yet. The final model was checked for correct material definitions and geometry errors until the OpenMC model of the VNS was confidently ensured.

\section{Simulation and Results}

\begin{figure}[ht]
\vspace{-6pt}
    \centering
    \begin{subfigure}[t]{0.24\textwidth}
        \centering
        \includegraphics[width=1.1\textwidth]{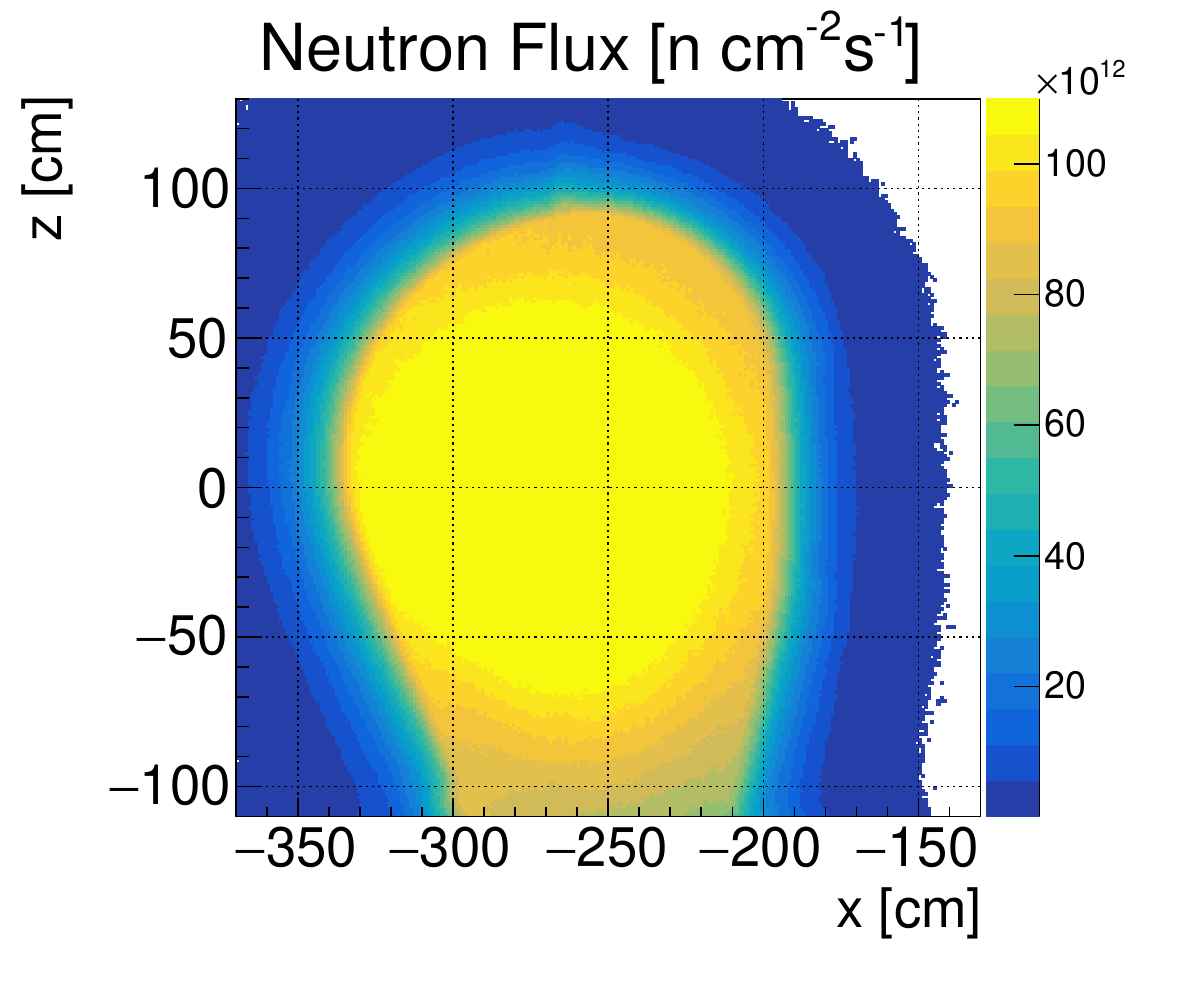}
        \caption{OpenMC neutron flux}
        \label{fig:xz_nflux}
    \end{subfigure}
    \hfill
    \begin{subfigure}[t]{0.24\textwidth}
        \centering
        \includegraphics[width=\textwidth]{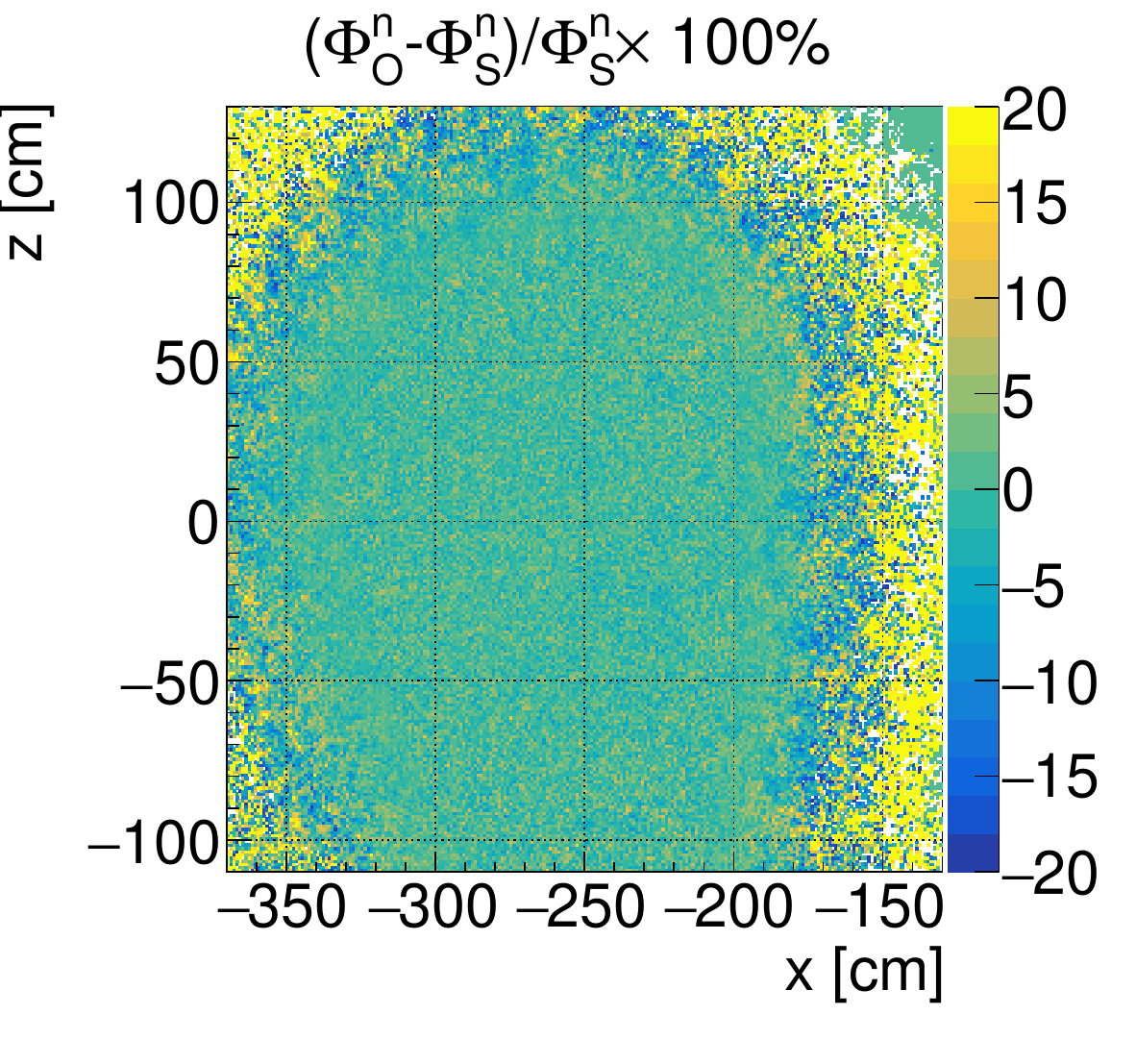}
        \caption{OpenMC-Serpent neutron flux relative difference}
        \label{fig:xz_nflux_ratio}
    \end{subfigure}
    \hfill
    \begin{subfigure}[t]{0.24\textwidth}
        \centering
        \includegraphics[width=\textwidth]{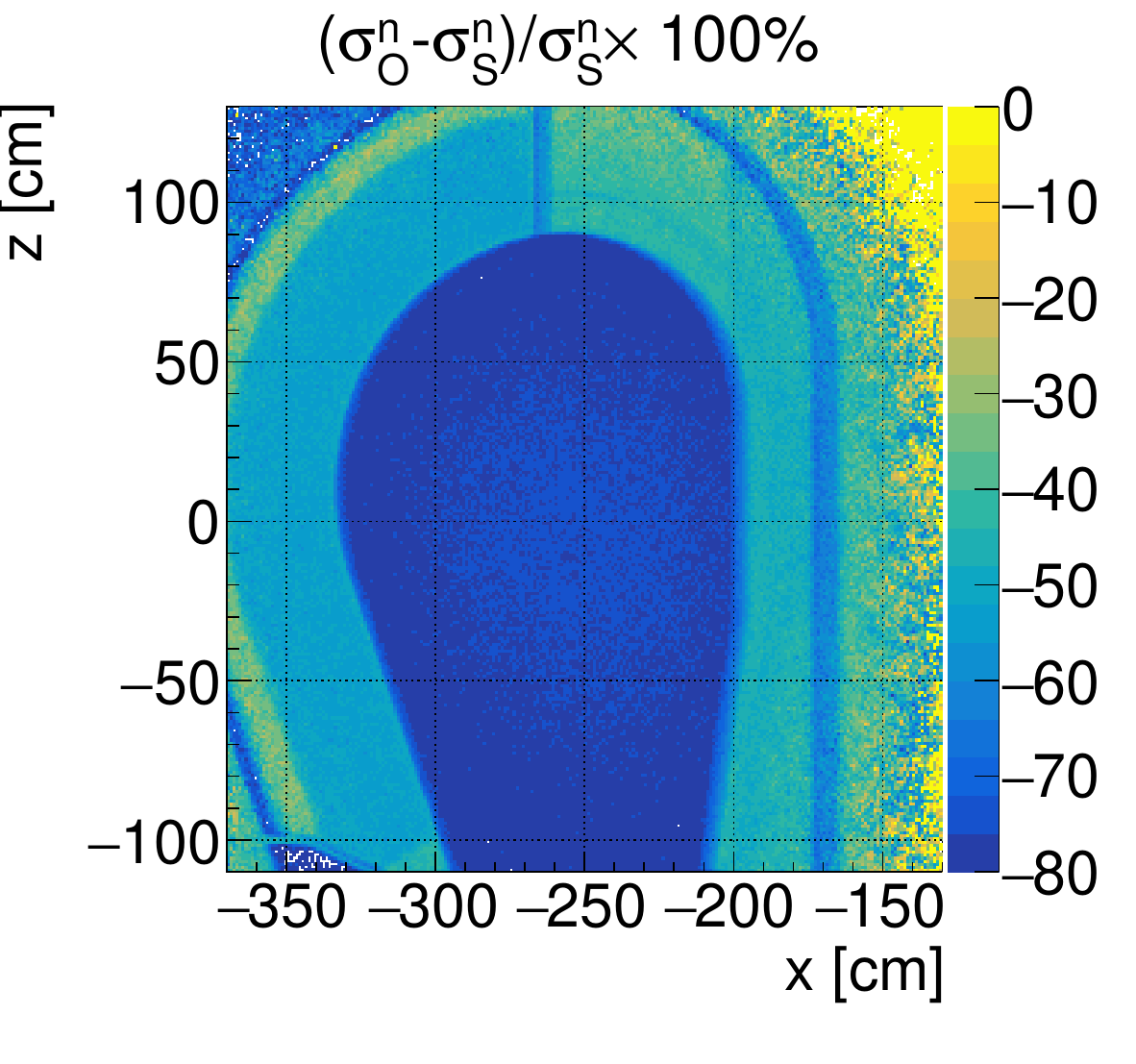}
        \caption{OpenMC-Serpent neutron flux standard error relative difference}
        \label{fig:xz_nflux_error}
    \end{subfigure}
    \begin{subfigure}[t]{0.24\textwidth}
        \centering
        \includegraphics[width=\textwidth]{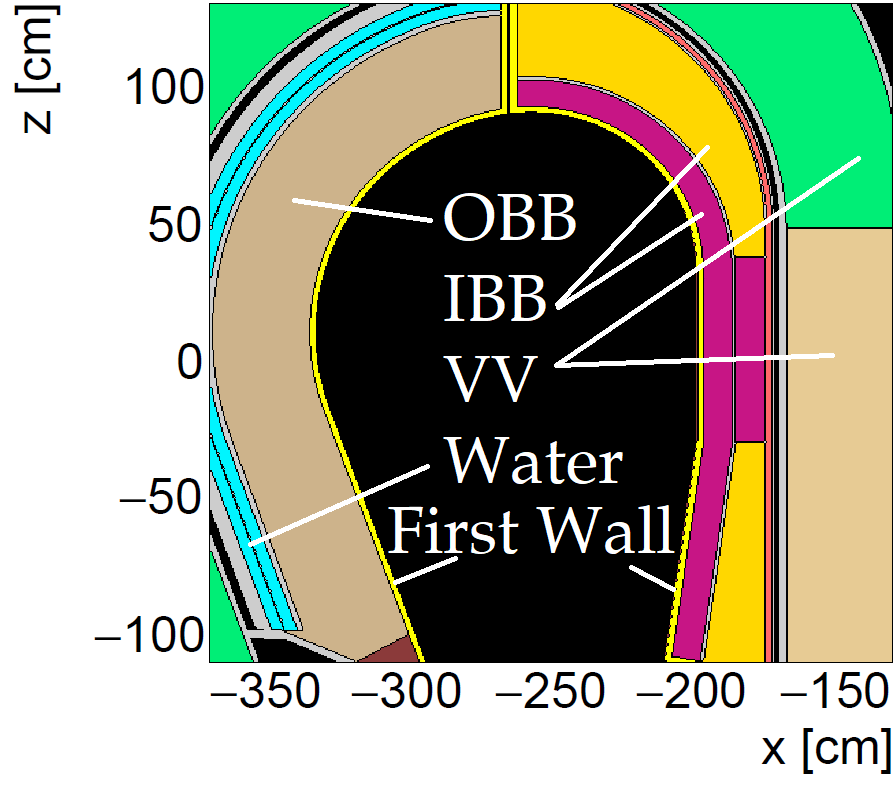}
        \caption{Geometry: In and out board blankets (IBB, OBB), vacuum vessel (VV), water, and first wall}
        \label{fig:xz_blanket_geometry}
    \end{subfigure}  
    \\
    \begin{subfigure}[t]{0.24\textwidth}
        \centering
        \includegraphics[width=1.1\textwidth]{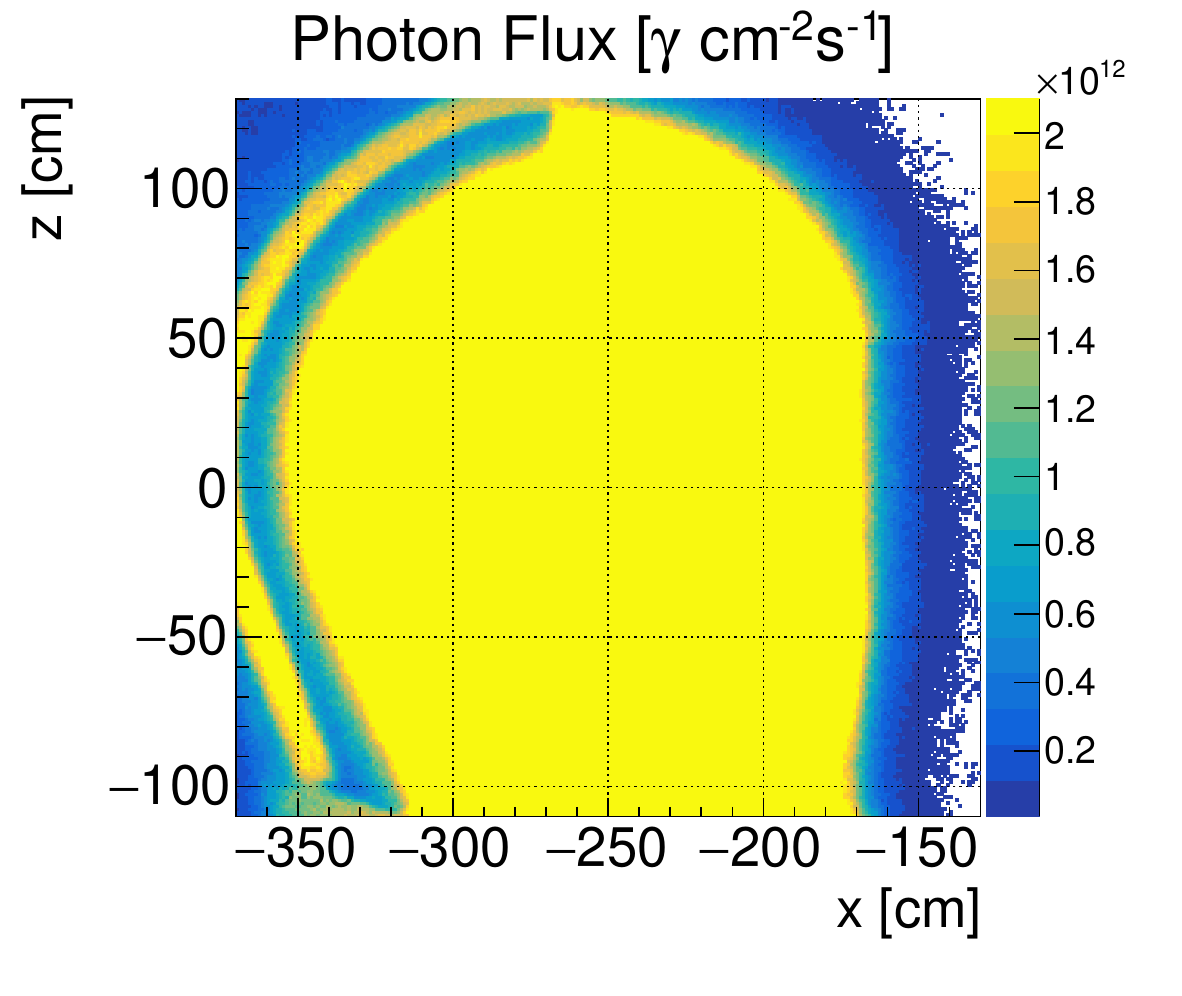}
        \caption{OpenMC photon flux}
        \label{fig:xz_pflux}
    \end{subfigure} 
    \hfill 
    \begin{subfigure}[t]{0.24\textwidth}
        \centering
        \includegraphics[width=\textwidth]{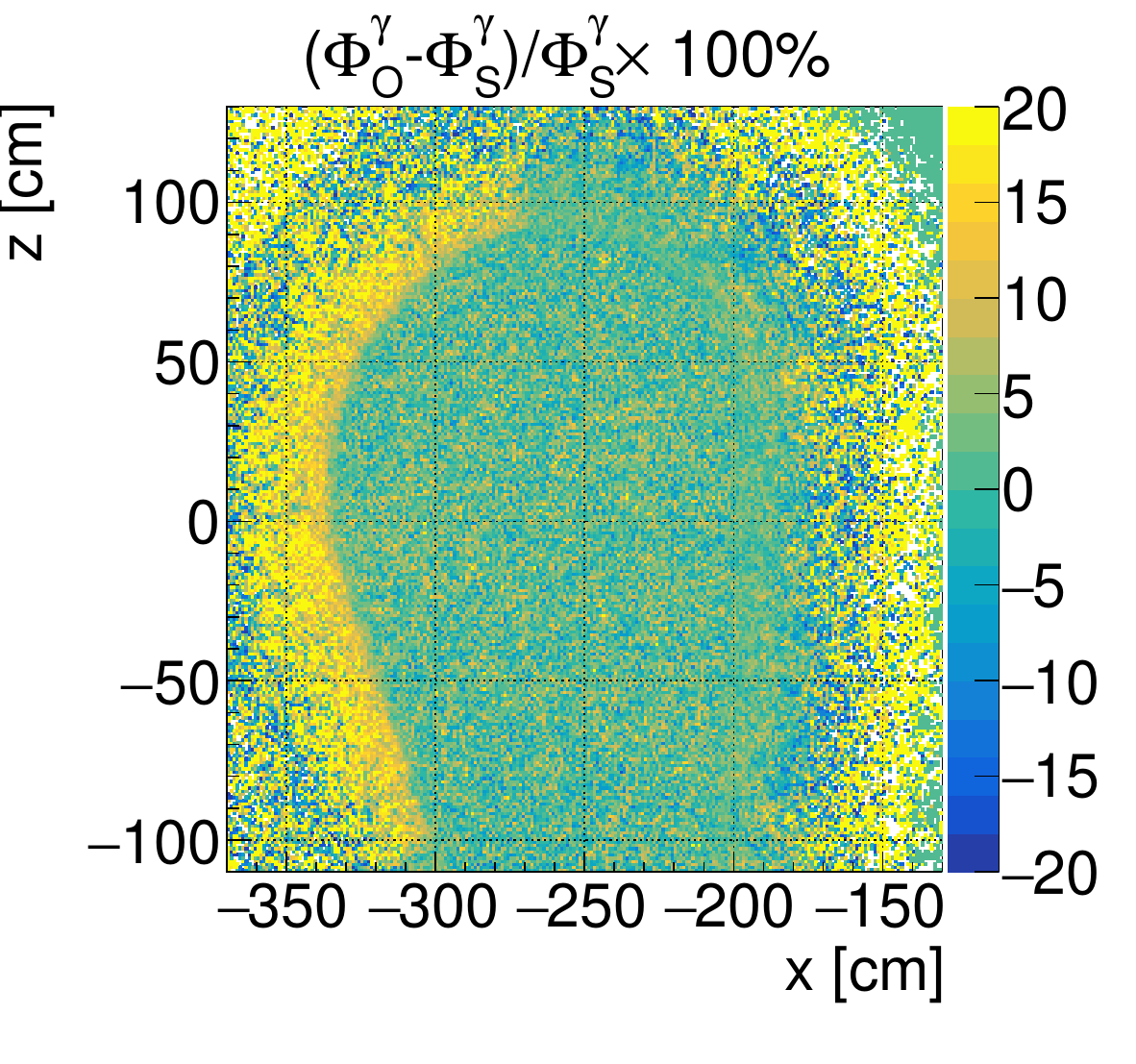}
        \caption{OpenMC-Serpent photon flux relative difference}
        \label{fig:xz_pflux_ratio}
    \end{subfigure}
    \hfill
    \begin{subfigure}[t]{0.24\textwidth}
        \centering
        \includegraphics[width=\textwidth]{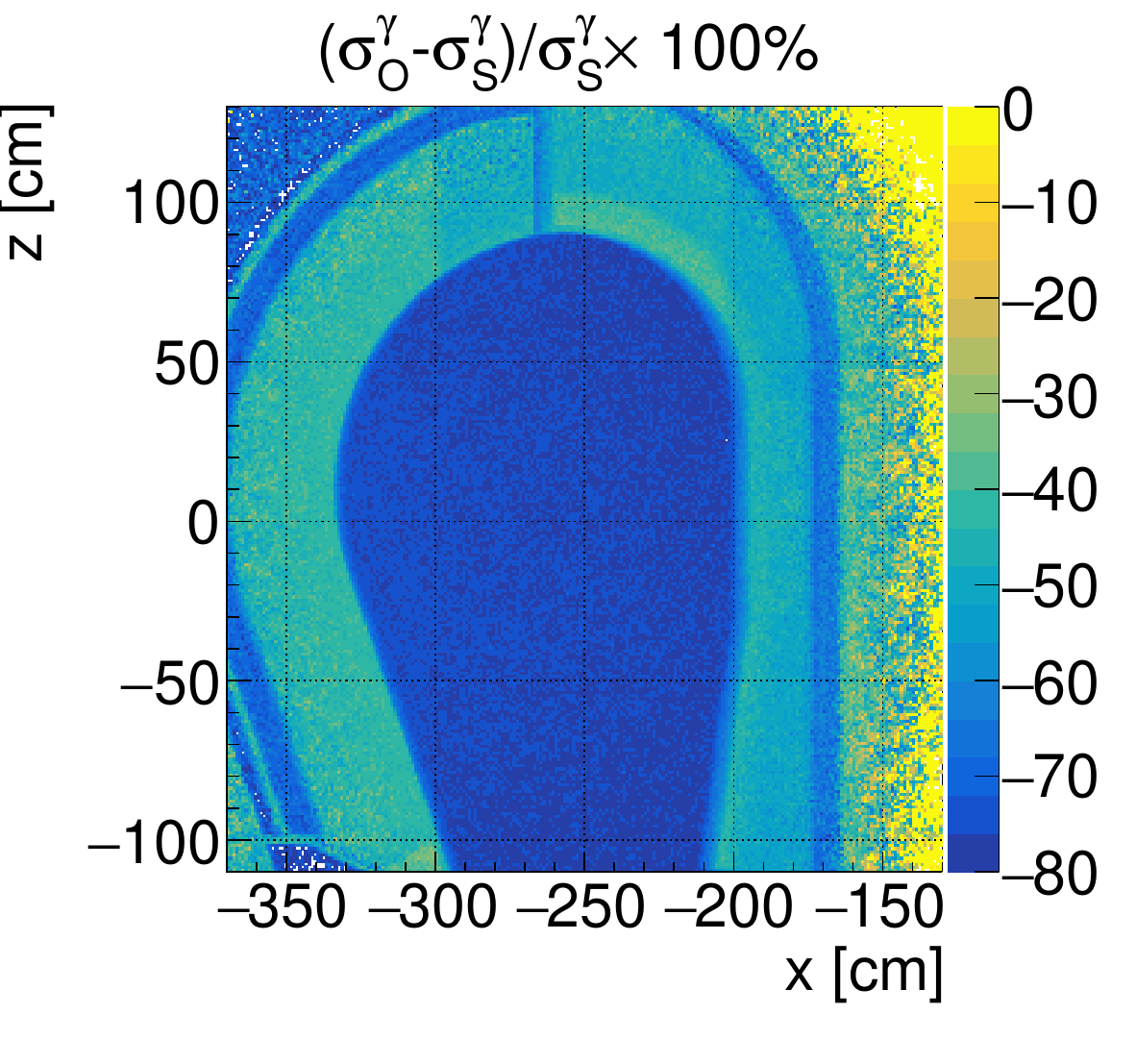}
        \caption{OpenMC-Serpent photon flux standard error relative difference}
        \label{fig:xz_pflux_error}
    \end{subfigure}
    \hfill
    \begin{subfigure}[t]{0.24\textwidth}
        \centering
        \includegraphics[width=\textwidth]{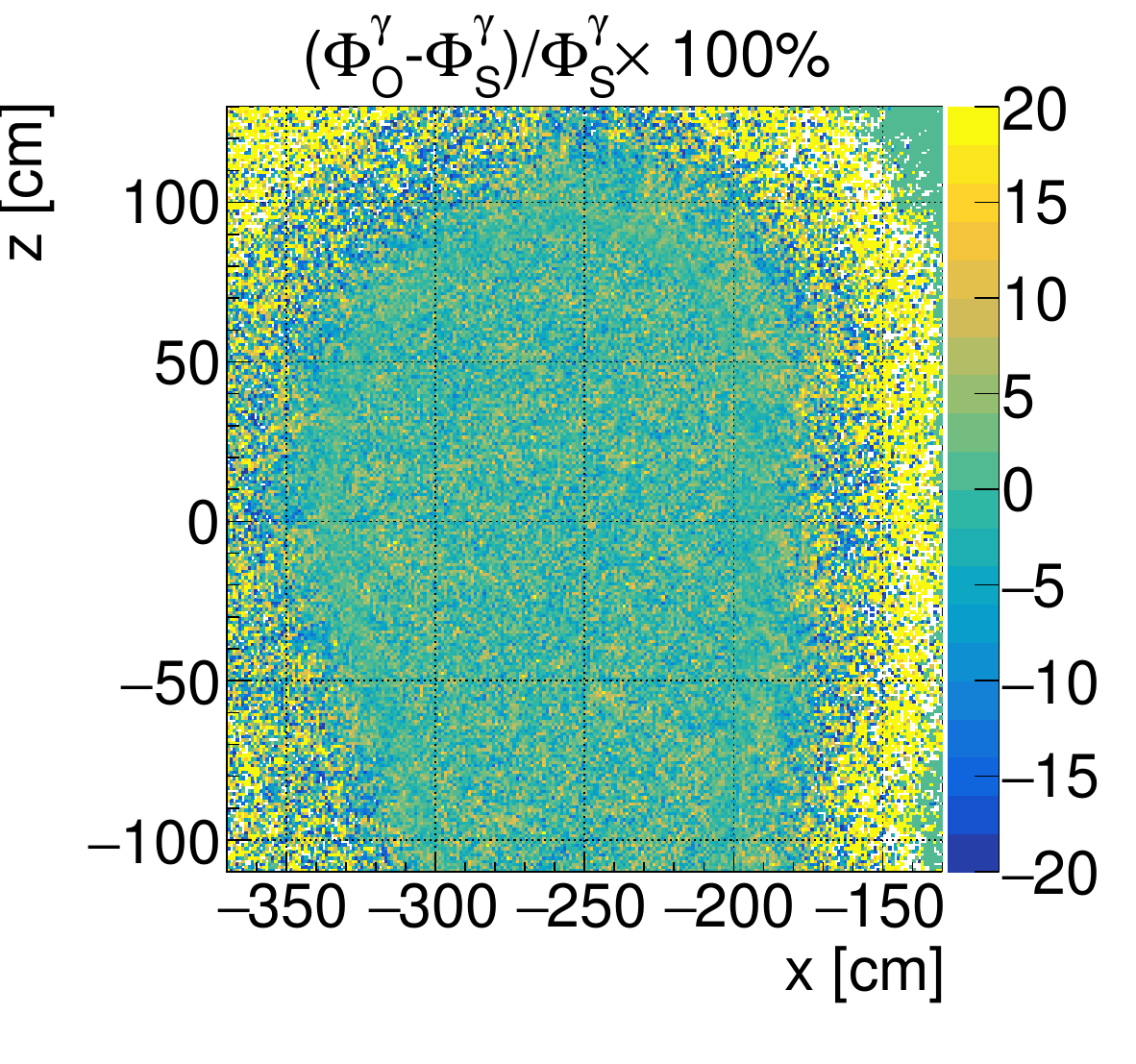}
        \caption{OpenMC-Serpent with delta tracking photon flux relative difference}
        \label{fig:xz_DTpflux_ratio}
    \end{subfigure}

    \label{fig:xz_flux}
    \caption{OpenMC and Serpent (``O'' and ``S'' subscripts) neutron and photon flux  in cut of vacuum vessel}
\vspace{-6pt}
\end{figure}

 The differences in VNS related calculations, were performed for a mapping of the neutron and gamma fluxes, $(n,T)$ and $(n,2n)$ reaction rates to verify in detail how sensitive the output results of Serpent and OpenMC are. Specific regions are inspected within the VNS to estimate possible discrepancies between codes for both reaction rates and particle flux values, as well as the quality of statistical uncertainty achieved in the calculations. The codes were each run with \SI{1e8}{neutron} histories divided into 1,000 batches in external source mode using the ENDF/B-VII.1 evaluated nuclear data library\cite{Chadwick20112887}. Neutron and photon flux detector responses compared the neutron flux of a cut through the VNS vacuum vessel (as show-cased in \cref{fig:xz_geometry} on the left side of the VNS). A geometry plot of the tallied region is shown in \cref{fig:xz_blanket_geometry}, including a cooling pipe, IB and OB blankets, the first wall, and \ce{TiH}$_2+$\ce{W} shielding, of the reactor for OpenMC and Serpent. The models resulted in very similar neutron and photon fluxes, agreeing to within -0.336\% and -1.52\%, suggesting slightly higher values in OpenMC than Serpent. The standard deviation values themselves were also compared (see \cref{fig:xz_nflux_error}), and it was found that the uncertainty was lower for OpenMC than Serpent, with the mean flux standard deviation relative difference being 67.4\% and photon flux being 83.0\% for neutrons and photons respectively. This difference is caused by the large void region, which suffers from lower statistics in Serpent and has to do with the detector response estimator used. The lower statistics for the photon flux results from the much lower photon population compared to neutrons (see \cref{fig:xz_nflux,fig:xz_pflux}).

Upon deeper inspection, the statistics are different between codes (see \cref{fig:xz_nflux_error}) despite cross-code agreement between neutron flux results in \cref{fig:xz_nflux_ratio}. This is very likely due to the different tally estimators across codes. OpenMC uses a track length estimator (TLE) to sample tallies, where as Serpent uses the collision flux estimator (CFE) for particles without interaction every \SI{20}{cm}, mixed with a track length estimator. The collision flux estimator depends on the collision frequency (and hence macroscopic cross section) of the material. \cref{fig:xz_nflux_error} and \ref{fig:xz_pflux_error} clearly shows that certain regions have higher statistics than others. The void region with no interaction probability, where neutrons are spawned, has about 75\% higher statistics in OpenMC's TLE tally collector compared to Serpent. The top left corner is another void region, where a higher statistic can be seen in OpenMC than Serpent, and just before that corner, is a region with much higher material density, and a more similar agreement in statistics can be seen between codes. 

As for photon flux direct response on the other hand, a difference on the order of $10\%$ to $20\%$ is observable in the outboard side. This is also likely due to the CFE used in Serpent, as the region is subject to strong changes in macroscopic cross section; the outboard blanket is surrounded by less dense regions both in front and behind it as is shown in the geometry plot \cref{fig:xz_blanket_geometry}. \cite{VALENTINE2022113197} also found a significant underestimation for photon flux in Serpent when compared with OpenMC.

\subsection{Reaction rate comparison}

\begin{figure}[ht]
    \centering
    \includegraphics[width=1.0\linewidth]{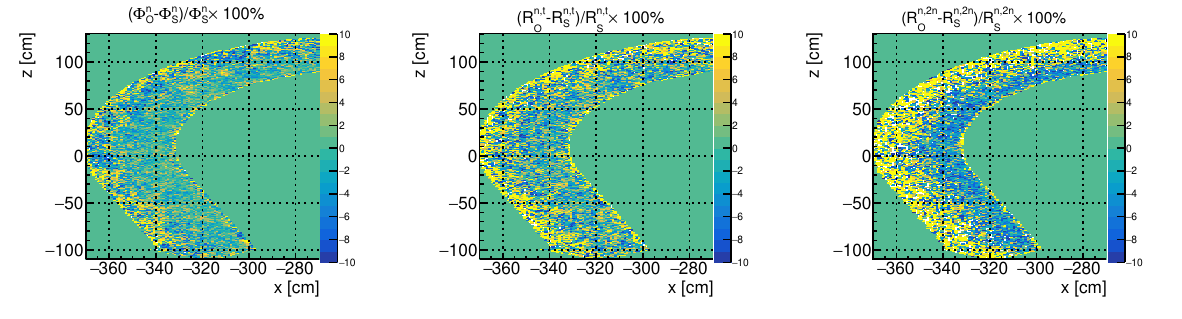}
    \caption{OB blanket OpenMC to Serpent relative difference of neutron flux, $(n,T)$, and $(n,2n)$ reacitons }
    \label{fig:OB_flux_ratio}
\vspace{-6pt}
\end{figure}
\begin{table}[ht]
  \centering
  \caption{OpenMC and Serpent with Hybrid and $\delta$ Tracking Estimator Total Outboard Blanket Flux, (\textit{n,T}) Reaction Rate, and (\textit{n,2n}) Reaction Rate}
  \label{tab:integral_reaction_rates}
  \begin{tabular}{lcccc}
    \toprule
    Code & $\Phi^n$ (n$\cdot $cm$^{-2}$s$^{-1}$) &$\Phi^\gamma$ ($\gamma\cdot$ cm$^{-2}$s$^{-1}$) & (\textit{n,T}) (s$^{-1}$)& (\textit{n,2n}) (s$^{-1}$)\\
    \toprule
    Serpent & $2.69\times10^{13}\pm 0.02\%$ & $4.27\times10^{12} \pm 0.02\% $ & $4.53\times10^{11}\pm 0.02\%$  & $1.27\times10^{11}\pm 0.02\%$ \\
    \midrule
    Serpent$_\delta$  & $2.69\times10^{13}\pm 0.02\%$ & $4.69\times10^{12} \pm 0.02\% $ & $4.52\times10^{11}\pm 0.02\%$  & $1.27\times10^{11}\pm 0.02\%$ \\
    \midrule
    OpenMC  & $2.65\times10^{13} \pm 0.02\%$& $4.69\times10^{12} \pm 0.02\%$ & $4.46\times10^{11} \pm 0.02\%$& $1.20\times10^{11} \pm 0.02\%$\\
    \midrule
    $\frac{OpenMC-Serpent}{Serpent}$ &  $-1.59\% \pm 0.02$ & $ 9.78\% \pm 2.00\%$ & $-1.55\% \pm 0.03\%$ & $-5.41\% \pm 0.03\%$\\
    \midrule
    $\frac{OpenMC-Serpent_{\delta}}{Serpent_{\delta}}$ &  $-1.56\% \pm 0.023\%$ & $-0.13\% \pm 0.03\%$ & $ -1.52\% \pm 0.03\%$ & $-5.38\% \pm 0.03\%$\\
    \bottomrule
  \end{tabular}
\end{table}
In addition to the triton production reaction, neutron multiplication reactions i.e. $(n,2n)$, is important to compensate a neutron flux depletion due to the absorption process in the blanket and to ensure sufficient Tritium breeding performance. These two reactions were tracked in the outboard blanket and compared as a ratio (shown in \cref{fig:OB_flux_ratio} with whole system integral results shown in \cref{tab:integral_reaction_rates}) and come to close agreement. Serpent systematically underestimates the tally results, with the larger reaction rate discrepancy coming from the $(n,2n)$, and furthermore OpenMC has a $9.78\%$ higher estimation for photon flux tallies than Serpent. The strong discrepancy in photon flux from \cref{fig:xz_pflux_ratio} is also reflected in the OB blanket integral value, which is not far off the $10-20\%$ discrepant OB blanket region. Two different tally estimators are used by default, Serpent uses a collision flux estimator (CFE), which samples virtual collisions every \SI{20}{cm} in addition to physical collisions, which is effective in low mean-free-path regions \cite{LEPPANEN2017161}. OpenMC uses the more versatile track length estimator (TLE), which records the length traveled by a particle between surface crossings and uses the region's macroscopic cross section to estimate reaction rates. As the geometry simulated is a mix of vacuum, and medium to strong absorbers, the low macrosopic cross sections regions may bias the simulation in CFE cases. Increasing the collision frequency by a factor of 10 was tested to see if more virtual collisions could yield closer results, but it did not. The Serpent simulation was rerun using delta tracking instead of hybrid delta-surface tracking, which uses surface tracking in regions with efficiency below $90\%$ real collisions. The efficiency was very low ($3.9\%$ for photon and $1.8\%$ for neutron) and yielded much closer results for photons (see \cref{fig:xz_DTpflux_ratio}), with only a $-0.13\%$ difference (with Serpent calculating a slightly higher photon flux). Employment of delta tracking cost time however, leading to $48\%$ longer computational time likely due to low sampling efficiency. The better agreement is likely due to virtual collision sampling in the strong absorbing OB blanket.
\subsection{Spectral flux comparison}
\begin{figure}[ht]
\vspace{-3pt}
    \centering
    \begin{subfigure}[t]{0.24\textwidth}
        \centering
        \includegraphics[width=1.25\textwidth]{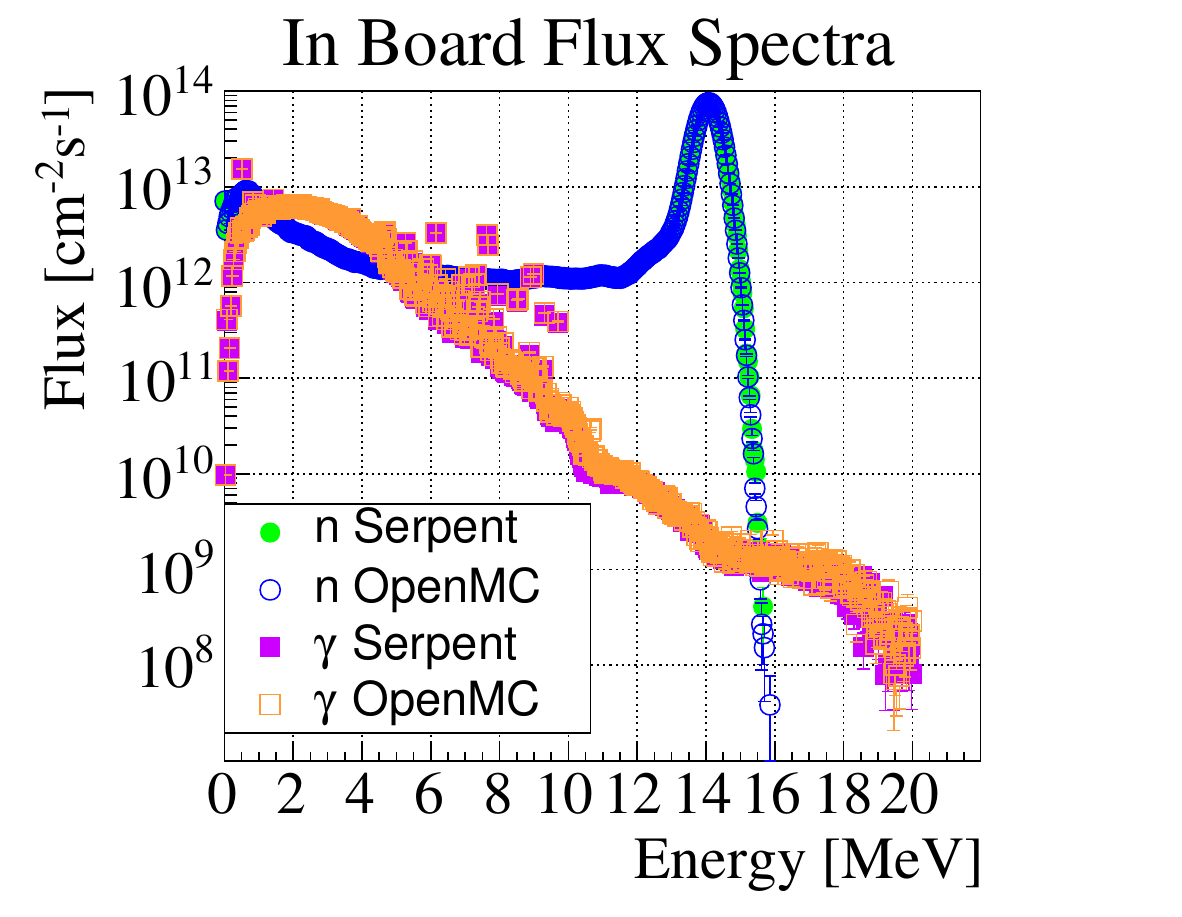}
        \caption{IB blanket Neutron and Photon flux spectra}
        \label{fig:IB_flux_spectra}
    \end{subfigure}
    \hfill
    \begin{subfigure}[t]{0.24\textwidth}
        \centering
        \includegraphics[width=1.25\textwidth]{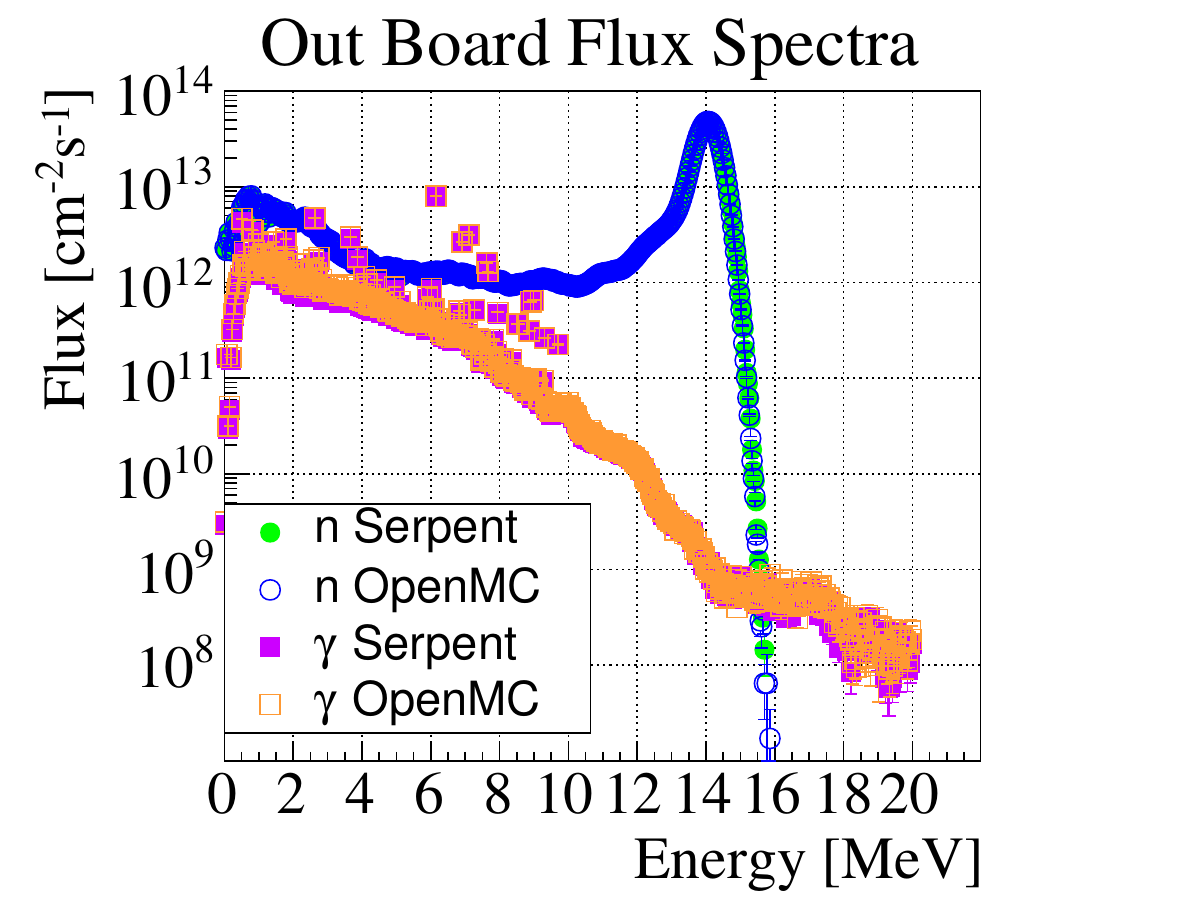}
        \caption{OB blanket Neutron and Photon flux spectra}
        \label{fig:OB_flux_spectra}
    \end{subfigure}
    \hfill
    \begin{subfigure}[t]{0.24\textwidth}
        \centering
        \includegraphics[width=1.25\textwidth]{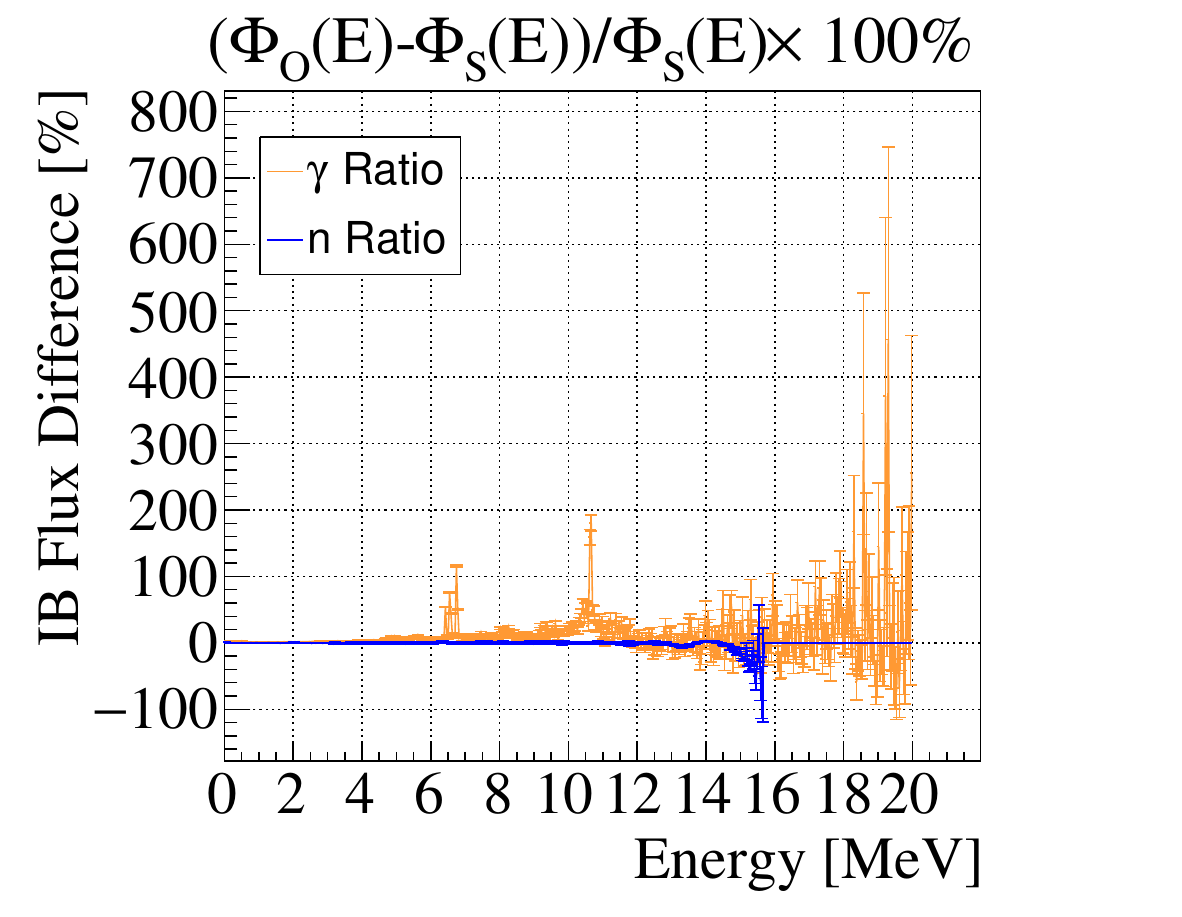}
        \caption{IB blanket Neutron and Photon flux spectra ratios}
        \label{fig:IB_flux_spectra_ratio}
    \end{subfigure}
    \hfill
    \begin{subfigure}[t]{0.24\textwidth}
        \centering
        \includegraphics[width=1.25\textwidth]{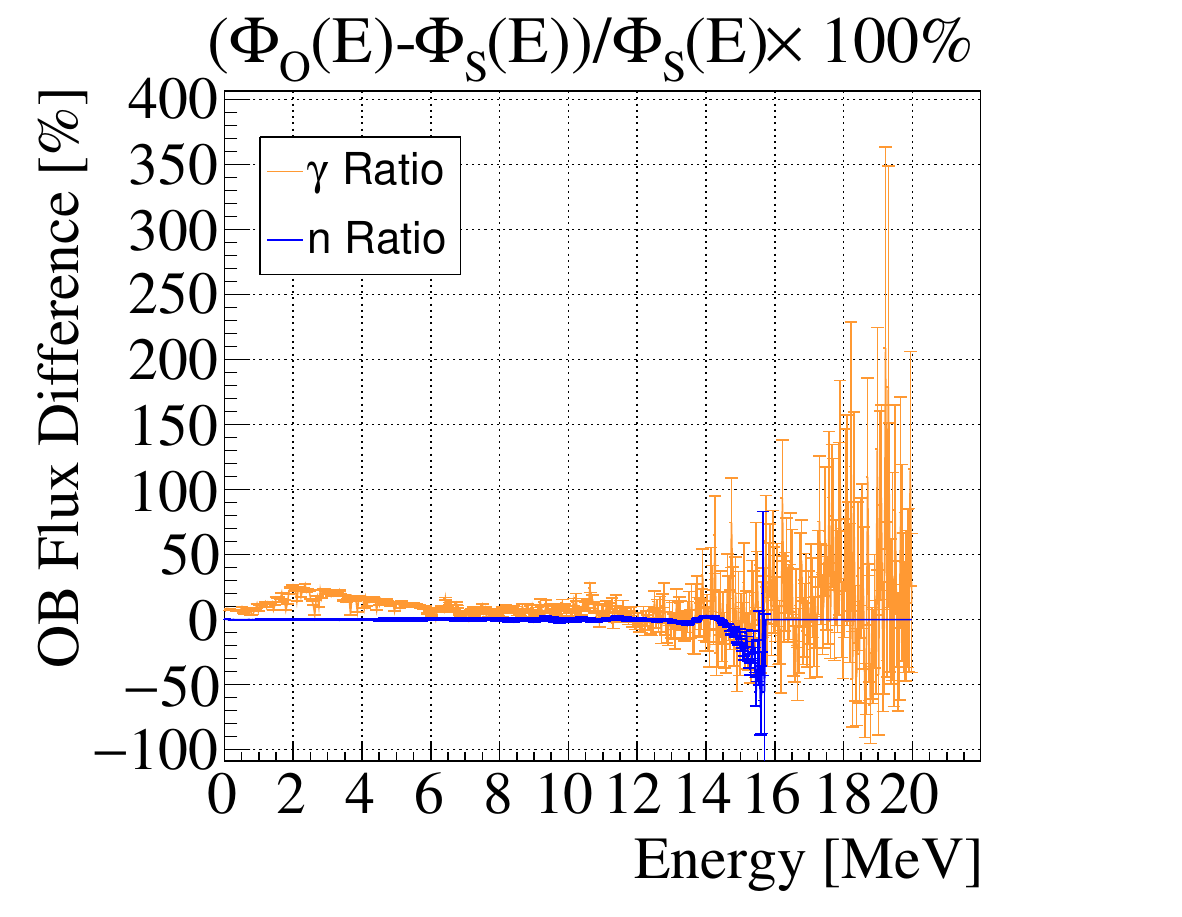}
        \caption{OB blanket Neutron and Photon flux spectra ratios}
        \label{fig:OB_flux_spectra_ratio}
    \end{subfigure}
    \label{fig:flux_spectra}
    \caption{Flux spectra comparison in blanket regions}
\vspace{-9pt}
\end{figure}
The neutron and photon flux of IB and OB blanket regions were plotted for OpenMC and Serpent in \cref{fig:IB_flux_spectra,fig:OB_flux_spectra}. Each detector uses 500 evenly spaced energy bins, from 0 to \SI{20}{MeV} for neutrons, and from \SI{7.5}{keV} to \SI{20}{MeV} for photons. In each blanket, except for in the high energy photon, and to a lesser extent neutron regimes, which suffer from low statistics, overall good agreement can be seen. One can see the \SI{14}{MeV} Gaussian overlaps nearly perfectly, justifying the source energy distribution used. In the high energy, low flux, low statistics tail of the distribution, slight overestimation comparable with the high error is seen by Serpent, increasing with energy. Individual resonances $<$\SI{10}{MeV} in the spectra visibly matching. It can be seen from \cref{fig:OB_flux_spectra_ratio} that the OpenMC OB photon flux spectrum has up to $\approx 25\%$ higher flux in the regime ($<$\SI{6}{MeV}), considering the discrepancy observed in \cref{tab:integral_reaction_rates} and \cref{fig:xz_pflux_ratio}. The IB flux discrepancies occur in lower flux regime ($>$\SI{10}{MeV}).
\subsection{Simulation time comparison}
\begin{minipage}{0.6\textwidth}
A comparison between total simulation times of the codes, in photon-neutron coupled simulation mode on the Leibniz Rechenzentrum HPC was performed by scaling up the number of nodes used. Each node is composed of a 160 Intel Xeon Platinum 8380 Ice Lake CPU threads. In each case, Serpent (using default particle tracking) was between 1.6 and $1.7\times$ faster than OpenMC, as plotted in \cref{fig:sim_time_upscale}. This result is likely due to the speed advantage of the CFE which samples fewer tallies than the TLE employed by OpenMC. This result is consistent with \cite{VALENTINE2022113197} which also presents Serpent results as faster than OpenMC for neutron-photon coupled simulation modes. When Serpent was used with full delta tracking, it slowed
\end{minipage}\hfill
\begin{minipage}{0.4\textwidth}
    \includegraphics[width=0.85\textwidth]{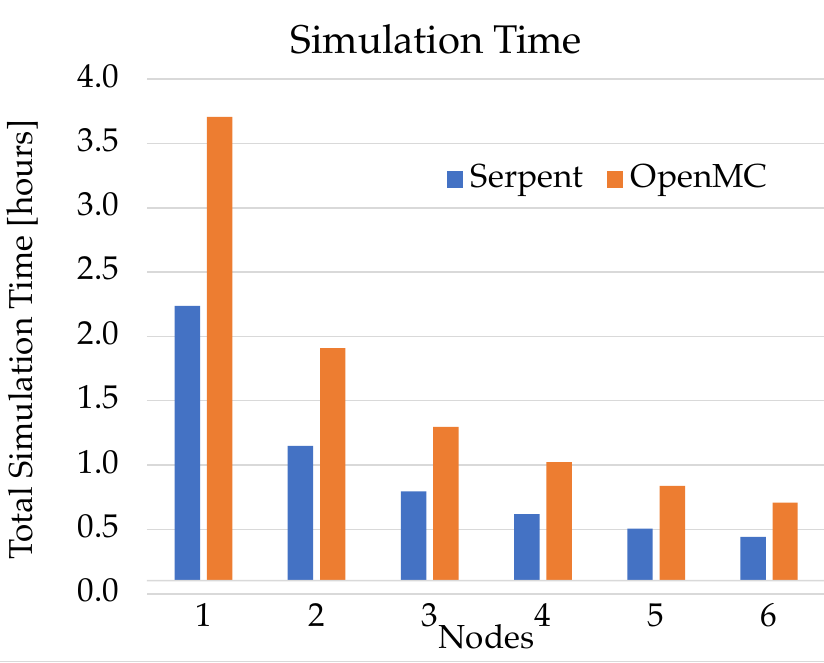}           
    \captionof{figure}{Simulation times on HPC}
    \label{fig:sim_time_upscale}
\end{minipage}
down to a similar computational time as OpenMC, where it only ran $1.12\times$ faster. Further comparison to \cite{VALENTINE2022113197} was made using neutron only mode, where \cite{VALENTINE2022113197} found  $1.85\times$ faster results in OpenMC. In VNS geometry, neutron only performance was $1.6\times$ faster in OpenMC than Serpent on one node. The slightly slower relative run time of OpenMC calculated in this work may be due to compiler issues of the OpenMC installation.
\subsection{Medical isotope generation}
The aim of this work is also to demonstrate confidence in the VNS model to simulate specific isotope production, making use of the advantages of a large scale fusion neutron source. A demonstrational production calculation was performed for different pathways of $^{99}$\ce{Mo} ($t_{1/2}=$ \SI{66}{hr}) generation, which decays into the medical isotope $^{99m}$\ce{Tc}. The VNS Serpent model was modified to arrange a dedicated space behind the first wall of the IB blanket. This space is reserved to accommodate thin tubes of \SI{2}{cm} diameter filled with capsules containing precursor materials. In both cases a total loading of \SI{97}{kg} of precursor material (depleted \ce{UO}$_2$ and $^{100}$Mo cases) has been included in the model. The IB blankets were chosen for the simulations of the medical isotope generation because: 1) the space and respectively shielding capacity of the blanket lost for the arrangement of the tubes do not noticeably affect the protection of the superconducting magnets behind it and 2) neutron flux is at maximum in this location. Burnup simulations using the predictor corrector method with half-day incremental burnup steps, during a 7-day irradiation \cite{pereslavtsev2024potential} were carried out using $10^{8}$\SI{}{neutron} histories, analog neutron-photon coupling mode, and default CFE tracking. The results discussed above enable us to apply Serpent code to produce high fidelity results for the neutron flux and spectra. Two different routes for $^{99}$\ce{Mo} production were simulated: 1)  the production via depleted \ce{UO}$_2(n,f)^{99}$\ce{Mo} and $^{100}$\ce{Mo}$(n,2n)^{99}$\ce{Mo} (complimented with the $^{100}$\ce{Mo}$(\gamma,n)^{99}$\ce{Mo} reaction) production channels. After a seven day irradiation followed by six days of decay (the typical $^{99}$\ce{Mo} processing plus delivery time), the activity yields are 3.3 six-day TBq $^{99}$\ce{Mo} and  1,300 six-day TBq $^{99}$\ce{Mo} for depleted \ce{UO}$_2$ and $^{100}$\ce{Mo} respectively. For a comparison, the Belgian research reactor BR-2, which accounted for $21\%$ of global $^{99}$\ce{Mo} production in 2016, has a production capacity of 290 six-day TBq/week \cite{NEA2016}.
\section{CONCLUSIONS}
Overall good agreement has been found for neutron transport between OpenMC and Serpent models of the VNS. OpenMC estimated higher tallies than Serpent, which could be due to the estimators used, as the Serpent model used the collision flux estimator, and for large distances the track length estimator. The (\textit{n,2n}) reaction had the highest difference for a reaction rate, with OpenMC predicting a 5.83\% higher reaction rate than Serpent. The photon flux tally had an even higher discrepancy, on the order of 10\% and higher, again with OpenMC reporting larger values. Upon further inspection, it was found that this too was a result from the Serpent CFE and agreement with OpenMC was found using Serpent's delta tracking functionality but at a cost of $48\%$ longer runtime. For default tracking mechanisms (CFE in Serpent, track length estimator in OpenMC) simulation time was between 1.6 and $1.7\times$ faster in Serpent than OpenMC in coupled neutron-photon transport mode, without any strong indication that further up-scaling nodes would lead to computational times faster than Serpent, although OpenMC is $1.6\times$ faster in neutron only mode. Two preliminary $^{99}$\ce{Mo} production routes were simulated and the results obtained indicate potential for the VNS as a medical isotope producer supplier.

\section*{ACKNOWLEDGEMENTS}
This work has been carried out within the framework of the EUROfusion Consortium, funded by the European Union via the Euratom Research and Training Programme (Grant Agreement No 101052200 — EUROfusion). Views and opinions expressed are however those of the author(s) only and do not necessarily reflect those of the European Union or the European Commission. Neither the European Union nor the European Commission can be held responsible for them. The authors gratefully acknowledge the computational and data resources provided by the Leibniz Supercomputing Centre (www.lrz.de). The authors would like to thank colleagues Gabriele Burgio and Daniel Bonete Wiese for their support during the preparation of this work.

\bibliographystyle{unsrt}
\bibliography{main}

\end{document}